\documentclass[12pt]{article}

\begin{document}

\begin{titlepage}

\begin{flushright}
IUHET-474\\
\end{flushright}
\vskip 2.5cm

\begin{center}
{\Large \bf Asymptotically Free Lorentz- and \\
CPT-Violating Scalar Field Theories}
\end{center}

\vspace{1ex}

\begin{center}
{\large B. Altschul\footnote{{\tt baltschu@indiana.edu}}}

\vspace{5mm}
{\sl Department of Physics} \\
{\sl Indiana University} \\
{\sl Bloomington, IN 47405 USA} \\

\end{center}

\vspace{2.5ex}

\medskip

\centerline {\bf Abstract}

\bigskip

We search for novel Lorentz- and CPT-violating field theories, beyond those
contained in the superficially renormalizable standard model extension. We find
a new class of scalar field self-interactions which are nonpolynomial in form,
involving arbitrarily high powers of the field. Many of these interactions
correspond to nontrivial asymptotically free theories. These theories are
stable if rotation invariance remains unbroken. These results indicate that
certain forms of Lorentz violation, if they exist, may naturally be quite
strong.

\bigskip

\end{titlepage}

\newpage

Recently, the possibility of there existing small Lorentz-violating corrections
to the standard model has received a great deal of attention, and there are now
numerous
experimental constraints limiting the magnitude of Lorentz-violating effects.
While there has also been a great deal of theoretical work in this area,
most systematic analyses of corrections to the standard model have focused only
on the superficially renormalizable Lorentz-violating effective field theories
whose general structure is described in~\cite{ref-kost1,ref-kost2,ref-kost3}.

We shall adopt a different point of view. In Lorentz-invariant scalar
field theories, there are known to exist interactions, which are nonpolynomial
in form (and thus superficially nonrenormalizable), yet actually are
renormalizable when the coupling constants' natural cutoff dependences are
taken into account~\cite{ref-huang1,ref-huang2}. There are
directions in the
parameter space of interactions in which the Gaussian fixed point is ultraviolet
(UV) stable,
and these correspond to nontrivial asymptotically free theories. As we shall
show, there are also
Lorentz-violating analogues of these nonpolynomial theories.

We shall examine the renormalization group (RG) flow for a class of
Lorentz- and CPT-violating scalar field theories. Our results should be relevant
to the study of Lorentz violation in the Higgs sector the standard model. Since
Lorentz violation in nature is a small effect, we shall restrict our attention
to the linearized form of the RG transformation, in which only terms that are
first-order in the Lorentz-violating interaction are retained. We shall use
the Wilsonian formulation of the RG~\cite{ref-wilson}, in which the theory is
considered with a momentum cutoff. Since small violations of Lorentz symmetry
may arise in a low-energy effective field
theory as remnants of larger violations appearing in a fundamental theory at
higher energies, it is very natural to study Lorentz violation in the context of
an effective theory with a cutoff.

In a previous paper~\cite{ref-altschul}, we analyzed the effects of Lorentz
violation on the known Lorentz-invariant nonpolynomial relevant potentials. We
shall follow the same method used in that paper here. The idea is to derive
a differential equation relating the form of a particular interaction with that
interactions' RG behavior. This strategy was first introduced
in~\cite{ref-periwal}, in which more sophisticated exact RG
techniques~\cite{ref-polchinski} were used to generalize the results
of~\cite{ref-huang1,ref-huang2}.

We shall be considering complex scalar field theories in which the current
operator
$j^{\mu}=i\left[\phi^{*}\left(\partial^{\mu}\phi\right)-\left(\partial^{\mu}
\phi^{*}\right)\phi\right]$ is contracted with a fixed vector $a^{\mu}$. The
simplest theory of this sort has Lagrange density
\begin{equation}
\label{eq-L0}
{\cal L}=\left(\partial^{\mu}\phi^{*}\right)\left(\partial_{\mu}\phi\right)+
a^{\mu}j_{\mu}-m^{2}\phi^{*}\phi.
\end{equation}
$a^{\mu}$ is an externally prescribed vector and is the source of the Lorentz
and CPT violation.
The action in this theory is a bilinear function of $\phi^{*}$ and $\phi$, and
so the theory is free. In particular, a field redefinition
\begin{equation}
\label{eq-atrans}
\phi\rightarrow e^{ia\cdot x}\phi,\,\phi^{*}\rightarrow e^{-ia\cdot x}\phi^{*}
\end{equation}
converts the Lagrange density into
\begin{equation}
{\cal L}'=\left(\partial^{\mu}\phi^{*}\right)\left(\partial_{\mu}\phi\right)
-(m^{2}+a^{2})\phi^{*}\phi.
\end{equation}
We see that the theory possesses a physical spectrum only for $a^{2}\geq-m^{2}$;
if $a^{2}<-m^{2}$, then the energy is not bounded below. When we consider more
general Lorentz-violating theories, the question of whether there is a positive
definite energy will continue to be very important, and we shall find that in
the interacting theories, much more stringent requirements on $a^{\mu}$ will be
needed in order to ensure stability.

To study the stability of a theory, we must work in Minkowski spacetime, in
which the energy is distinguished from the other components of the
four-momentum. However, it will be much simpler for us to perform our
RG calculations in Euclidean space. We may then transform our results back into
Minkowski spacetime and determine whether or not the energy is bounded below.


The gauge symmetries of the standard model
restrict the allowed forms of any Lorentz-violating Higgs
couplings~\cite{ref-kost1,ref-kost5}. There is only one such gauge invariant,
superficially renormalizable, CPT-violating interaction---a
generalization of (\ref{eq-L0}), with a single value of $a^{\mu}$ for both
field components. Any nonpolynomial interactions are similarly
restricted, so, while we shall consider only the case of a single
complex scalar field, a generalization of our results should be relevant to the
study of the physical Higgs sector.

We shall use bare perturbation theory to calculate the lowest-order radiative
corrections to the effective action. Our calculations will be applicable in any
number of dimensions $d>2$, although obviously $d=4$ is the most important.
The Euclidean action for the bare theory is
\begin{equation}
S=\int d^{d}x\left\{\left(\partial_{j}\phi^{*}\right)\left(\partial_{j}\phi
\right)-m^{2}\phi^{*}\phi+ia_{j}\left[\phi^{*}\left(\partial_{j}\phi\right)-
\left(\partial_{j}\phi^{*}\right)\phi\right]V\left(\phi^{*}\phi\right)\right\}.
\end{equation}
The function $V$ parameterizing the interaction must be representable as a power
series in $|\phi|^{2}$. When $V(\phi^{*}\phi)$ is not constant, the interaction
term cannot be eliminated by means of the transformation (\ref{eq-atrans}).
We shall analyze the RG flow for
this action, utilizing the same techniques as were used in~\cite{ref-altschul}.
This shall consist of determining the effective $n$-point vertex for the
theory, by summing up the contributions of an infinite number of tadpole
diagrams, each involving
a single bare $(n+2k)$-point vertex and $k$ loops.

We shall regulate this theory with a momentum cutoff $\Lambda$, which will
provide the only intrinsic scale in the theory. The coupling constants'
classical dependences on $\Lambda$ will therefore be determined entirely by
their dimension. A coupling $g_{K}$ with dimension (mass)$^{d_{K}}$ will be
associated with a dimensionless coupling constant $c_{K}$ according to
\begin{equation}
\label{eq-uK}
g_{K}=c_{K}\Lambda^{d_{K}}.
\end{equation}
The extra factors of $\Lambda$ in (\ref{eq-uK}) ensure the renormalizability of
all interactions, essentially because any superficially nonrenormalizable
couplings will vanish
as $\Lambda\rightarrow\infty$. However, the dimensionless couplings $c_{K}$ may
remain finite, and it is the evolution of the $c_{K}$ under the action of
the RG that is important.
The inclusion of the extra factors of $\Lambda$ in (\ref{eq-uK}) will allow us
to write down a $\Lambda$-independent differential equation describing the
normal modes of the RG flow. The explicit representation of $V(\phi^{*}\phi)$
in terms of dimensionless couplings is given below, in (\ref{eq-Sint}) and
(\ref{eq-Usum}).

According to the scaling scheme just described, the mass parameter should have
the form $m^{2}=\mu^{2}\Lambda^{2}$. In
$\phi^{4}$ theory, which becomes trivial as $\Lambda\rightarrow\infty$, the
cutoff and mass scales are not generally well separated, and self-consistency
conditions relating the two scales have been used to put upper bounds on the
standard model's Higgs mass~\cite{ref-dashen,ref-kuti}. However, in the presence
of asymptotically free
interactions, there is no reason why the cutoff should not be arbitrarily large,
so it is possible in principle for $\mu^{2}$ to be very small.

We shall consider only the first term in the interaction: $ia_{j}\left[\phi^{*}
\left(\partial_{j}\phi\right)\right]V\left(\phi^{*}\phi\right)$. The
contributions from
the other term are only trivially different. A particular term in the power
series expansion of $V$ contributes
\begin{equation}
{\cal L}_{n}=ia_{j}\left[\phi^{*}\left(\partial_{j}\phi
\right)\right]\left(\phi^{*}\phi\right)^{n}
\end{equation}
to the Lagrange density. When the Feynman rules for the theory are worked out,
${\cal L}_{n}$ gives rise to a vertex with $n+1$ incoming particle lines
(corresponding to $\phi$) and an equal number of outgoing ($\phi^{*}$)
lines. Moreover, in addition to the various combinatorial factors
associated with the vertex, there is a factor of $\vec{a}\cdot\vec{p}$, where
$\vec{p}$ is the momentum on one of the incoming legs; we must sum over all
such incoming legs to which this momentum may be assigned.

The linearized RG flow is generated by diagrams with a single vertex of the form
generated by ${\cal L}_{n}$, with $k$ outgoing and $k$ incoming particle lines
connected to form $k$ tadpole loops. This generates an effective vertex, with a
corresponding effective Lagrange density of the form ${\cal L}_{n-k}$.
Determining the combinatorial factors associated with the effective vertex is a
relatively simple matter. If we begin with a ${\cal L}_{n}$-type diagram and
contract one pair of lines into a loop, we get a factor of $n+1$ arising from
the choice of which outgoing line is to be used and a factor of $n$
from the choice of the incoming line. The difference between the two factors
arises from the fact that we may only choose one of the incoming lines without
the extra factor of $\vec{a}\cdot\vec{p}$. If we did choose the leg with the
momentum factor attached, we would obtain a loop integral whose integrand was an
odd function of the momentum. Hence, the contribution from this contraction
would vanish. \{This argument also guarantees that there can be no ${\cal O}
\left[\left(\phi^{*}\phi\right)^{0}\right]$ contribution to the effective action
generated by the tadpole loops; there must be at least one external leg on each
nonvanishing diagram, to which the momentum factor may be attached.\}

When acting on $\left(\phi^{*}\phi\right)^{n}$, the operator
\begin{equation}
{\cal D}=\frac{\partial^{2}}{\partial\left(\phi^{*}\phi\right)^{2}}\left(
\phi^{*}\phi\right)
\end{equation}
generates $n(n+1)\left(\phi^{*}\phi\right)^{n-1}$. So this differential operator
will produce the necessary combinatorial factors accompanying a loop, when it
acts
on $V\left(\phi^{*}\phi\right)$. Each loop is also associated with a factor of
$D_{F}(0)$, the Feynman propagator for the complex scalar field at zero spatial
separation. Moreover, a
diagram with $k$ loops has a symmetry factor of $k!$, because we are free to
interchange the loops.

The value of $D_{F}(0)$ is
\begin{equation}
D_{F}(0)=\int_{|p|<\Lambda}\frac{d^{d}p}{(2\pi)^{d}}\frac{1}{p^{2}+m^{2}}.
\end{equation}
We see that $D_{F}(0)$ has the form $C\Lambda^{d-2}$, with
\begin{equation}
\label{eq-C}
C=\frac{1}{\pi^{d}2^{d-1}\Gamma(d/2)}\int_{0}^{1}d\xi\,\frac{\xi^{d-1}}{\xi^{2}
+\mu^2}.
\end{equation}
For $d=4$, $C=\frac{1}{16\pi^{2}}\left[1-\mu^{2}\log\left(1+\mu^{-2}\right)
\right]$.
The integral in (\ref{eq-C}) may also be approximated by $\frac{1}{d-2}$ if
$\mu^{2}\ll 1$.
The $d-2$ factors of $\Lambda$ appearing
in the zero-separation propagator are the source of the RG flow.

In order to study this RG flow, we must express the interactions in the
nondimensionalized form described previously.
Recalling that we are neglecting the $(\partial\phi^{*})\phi$ term,
we therefore write the interaction part of the effective action as
\begin{equation}
\label{eq-Sint}
S_{int}=\int d^{d}x\,ia_{j}\left[\phi^{*}\left(\partial_{j}\phi
\right)\right]U\left[\Lambda^{-(d-2)/2}\phi^{*}\phi\right],
\end{equation}
Since the action is dimensionless and $ia_{j}\left[\phi^{*}\left(\partial_{j}
\phi\right)\right]$ has dimension (mass)$^d$, $U$ is a dimensionless function.
In addition to the explicit $\Lambda$-dependence in this expression,
$ia_{j}\left[\phi^{*}\left(\partial_{j}\phi\right)\right]$ scales as
$\Lambda^{d}$, and $U\left[\Lambda^{-(d-2)/2}\phi^{*}\phi\right]$ may
also have a parametric dependence upon $\Lambda$; this
parametric dependence will describe the RG flow of the effective potential.

If only the tadpole diagrams contribute, then
we may calculate $U\left[\Lambda^{-(d-2)/2}\phi^{*}\phi\right]$ directly from
$V\left(\phi^{*}\phi\right)$. Accounting for the combinatorial factors described
above, the result is that
\begin{eqnarray}
U\left[\Lambda^{-(d-2)/2}\phi^{*}\phi\right] & = & \sum_{k=0}^{\infty}\frac{1}
{k!}\left[-D_{F}(0){\cal D}\right]^{k}V\left(\phi^{*}\phi\right) \\
\label{eq-expD}
& = & \exp\left[-C\Lambda^{d-2}{\cal D}\right]V\left(\phi^{*}\phi\right).
\end{eqnarray}
Acting with the operator $\Lambda\frac{d}{d\Lambda}$ on
(\ref{eq-expD}) results in the differential equation
\begin{equation}
\label{eq-PDE}
\Lambda\frac{\partial U}{\partial\Lambda}-(d-2)
\left[\Lambda^{-(d-2)}\phi^{*}\phi\right]
U'\left[\Lambda^{-(d-2)}\phi^{*}
\phi\right]=-(d-2)C\Lambda^{d-2}{\cal D}U\left[\Lambda^{-(d-2)}\phi^{*}\phi
\right],
\end{equation}
where the prime denotes differentiation of $U$ with respect to its argument.
The left-hand side of (\ref{eq-PDE}) describes the classical scaling behavior
of $U$, while the right-hand side  contains the effects of quantum
corrections.

If $U$ is to describe a normal mode of the RG flow near the free-field
fixed point, then it should display
a power-law dependence on $\Lambda$. This is, we should have
$\Lambda\frac{\partial U}{\partial\Lambda}=-\lambda U$, for some constant
$\lambda$. In this case, the
partial differential equation (\ref{eq-PDE}) is transformed into the
ordinary differential equation
\begin{equation}
\label{eq-ODE}
\lambda U(y)+(d-2)yU'(y)-(d-2)C\left[yU''(y)+U'(y)\right]=0,
\end{equation}
where $y=\Lambda^{-(d-2)}\phi^{*}\phi$ is the argument of $U$.
To solve the equation (\ref{eq-ODE}), we set
\begin{equation}
\label{eq-Usum}
U=\sum_{n=0}^{\infty}c_{n}y^{n},
\end{equation}
and this leads to the recurrence relation
\begin{equation}
c_{n+1}=\frac{\lambda+(d-2)n}{(d-2)(n+1)(n+2)C}c_{n}.
\end{equation}
So the solution is
\begin{equation}
U(y)=gM\left(\frac{\lambda}{d-2};2;\frac{y}{C}\right),
\end{equation}
where $g$ is some coupling constant, and $M(\alpha;\beta;z)$ is the confluent
hypergeometric (Kummer) function~\cite{ref-abramowitz}
\begin{equation}
M(\alpha;\beta;z)=1+\frac{\alpha}{\beta}\frac{z}{1!}+\frac{\alpha(\alpha+1)}
{\beta(\beta+1)}\frac{z^{2}}{2!}+\cdots=\frac{\Gamma(\beta)}{\Gamma(\beta-
\alpha)\Gamma(\alpha)}\int_{0}^{1}dt\, e^{zt}t^{\alpha-1}(1-t)^{\beta-\alpha-1}.
\end{equation}
We see that $U(y)$ has a polynomial form exactly if $\frac{\lambda}{d-2}$ is a
nonpositive integer.
The full renormalized $S_{int}$ corresponding to this normal mode of the RG flow
[including the thus far omitted $(\partial\phi^{*})\phi$ term] is
\begin{equation}
S_{int}=g\int d^{d}x\,ia_{j}\left[\phi^{*}\left(\partial_{j}\phi\right)
-\left(\partial_{j}\phi^{*}\right)\phi\right]M\left[\frac{\lambda}{d-2};2;
\frac{\Lambda^{-(d-2)/2}\phi^{*}\phi}{C}\right].
\end{equation}
$\lambda$ is clearly the anomalous scaling dimension of the $U$ factor in
$S_{int}$. The operator $ia_{j}\left[\phi^{*}\left(\partial_{j}\phi-
(\partial_{j}\phi^{*})\phi\right)\right]$ 
has dimension (mass)$^{d}$, so the anomalous dimension of the
entire interaction is $\lambda+d$.
This scaling dimension describes how the interaction scales with changes in
the cutoff $\Lambda$; it need not describe the scaling of any correlation
functions with respect to their external momenta. The calculation of such
correlation functions involves the same sorts of complexities as are associated
with similar calculations in the presence of Lorentz-invariant nonpolynomial
potentials~\cite{ref-periwal,ref-halpern}.

If $\lambda>0$, the potential is nonpolynomial [with $U(y)$ behaving as
$y^{\lambda/(d-2)-2}e^{y/C}$ for large $y$] and asymptotically free, with
power-law coupling constant flow. The
Gaussian fixed point is UV stable along the associated trajectories; if the
renormalized coupling is held fixed, then the bare coupling goes to zero as
$\Lambda\rightarrow\infty$. If $\lambda<0$, then the fixed point is infrared
stable, and the interaction is irrelevant; this includes all interactions with
polynomial $U$. The marginal case, $\lambda=0$, corresponds to the free theory
discussed earlier.

We must now determine whether the relevant interactions we have found lead to
stable Minkowski-space theories. In fact, the canonical Hamiltonian density
associated with the Lagrange density
\begin{equation}
{\cal L}=\left(\partial^{\mu}\phi^{*}\right)\left(\partial_{\mu}\phi\right)+
ia^{\mu}\left[\phi^{*}\left(\partial_{\mu}\phi\right)-\left(\partial_{\mu}
\phi^{*}\right)\phi\right]V\left(\phi^{*}\phi\right)-m^{2}\phi^{*}\phi
\end{equation}
is
\begin{equation}
\label{eq-H}
{\cal H}=\left|\pi-ia^{0}\phi^{*}V\left(\phi^{*}\phi\right)\right|^{2}+
\left(\partial_{j}\phi^{*}\right)\left(\partial_{j}\phi\right)+m^{2}\phi^{*}\phi
+ia_{j}\left[\phi^{*}\left(\partial_{j}\phi\right)-\left(\partial^{\mu}
\phi^{*}\right)\phi\right]V\left(\phi^{*}\phi\right).
\end{equation}
All the terms in (\ref{eq-H}) are manifestly positive except for the last one.
If any spatial components of $a^{\mu}$ are nonvanishing
and $V$ grows more rapidly than a constant, then this
term can render ${\cal H}$ arbitrarily negative.
Hence we conclude that the theory must be quantized in a reference frame
in which $a^{\mu}$ is purely timelike (i.e. $a_{j}=0$ for $j=1,2,3$); such a
frame obviously can exist only if $a^{2}\geq0$. In this special quantization
frame, rotation invariance remains an unbroken symmetry, and the momentum
regulator is symmetric. If the theory is boosted into a different frame, then a
boosted regulator will be required.
It is well established
that consistency requirements in quantum field theories may constrain the
values of Lorentz-violating coefficients~\cite{ref-kost3}. Moreover, a purely
timelike $a^{\mu}$ is appealing from a physical standpoint, because the
universe displays a very high degree of isotropy in the reference frame of the
cosmic microwave background, and this limits the possibilities for spacelike
Lorentz-violating effects.

However, the existence of strongly relevant directions in the parameter space of
Lo\-rentz-violating interactions does present a sort of ``hierarchy problem." A
generic field
theory containing a scalar boson sector and Lorentz-violating coefficients with
the same discrete
symmetries as $a^{0}$ (C-odd, P- and T-even) will generate, through radiative
corrections, scalar self-interactions of the sort we have considered here. Since
some of these interactions are asymptotically free, we expect them to be
generated fairly strongly. The strong scalar field interactions will, in turn,
generate
Lorentz-violating interactions in other sectors of the theory.
However, since Higgs-sector
Lorentz violation in nature is a very weak effect~\cite{ref-anderson},
we must conclude that either there is some additional symmetry
that prevents the generation of strong $a^{0}$-type effects in the observable
sectors of the theory, that the bare Lorentz-violating couplings in
question are all extraordinarily small, or that nonlinear effects (possibly
involving other interactions) become important even for very small values of
$g$.
To study this problem further, it would be desirable if the calculation of the
RG flow could be extended beyond the linearized regime, but that could be a very
difficult undertaking.

In summary, we see that if Lorentz- and CPT-violating corrections to the
standard
model do exist, then there is good reason to believe that they may not be small,
since the Higgs sector can support asymptotically free Lorentz-violating
interactions. We have determined the specific forms that these interactions take
in the weak coupling limit and studied the stability of the corresponding
theories. The requirement that the coefficient $a^{\mu}$ parameterizing the
Lorentz violation be purely timelike arose quite naturally in our analysis. The
net result is that we have obtained important theoretical insights about
the forms to be taken by any possible Lorentz-violating interactions.

\section*{Acknowledgments}
The author is grateful to V. A. Kosteleck\'{y} and K. Huang for
their helpful discussions.
This work is supported in part by funds provided by the U. S.
Department of Energy (D.O.E.) under cooperative research agreement
DE-FG02-91ER40661.

\end{document}